\documentclass[12pt]{iopart}
\usepackage{iopams}
\usepackage{graphicx}
\usepackage{dcolumn}
\usepackage{bm}
\usepackage{amssymb}
\usepackage{verbatim}
\usepackage{epstopdf}
\usepackage{xcolor}
\usepackage{soul}

\newcommand{\beq}{\begin{equation}}
\newcommand{\eeq}{\end{equation}}
\newcommand{\bea}{\begin{eqnarray}}
\newcommand{\eea}{\end{eqnarray}}

\newcommand{\eq}[1]{{(\ref{#1})}}
\newcommand{\commentout}[1]{{}}

\newcommand{\half}{{\hbox{$\frac{1}{2}$}}}

\begin{document}
\title{Emergent classicality in continuous quantum measurements}
\author{Juha Javanainen}
\address{Department of Physics, University of Connecticut, Storrs, Connecticut 06269-3046, USA}
\author{Janne Ruostekoski}
\address{School of Mathematics, University of Southampton,
Southampton, SO17 1BJ, UK}
\ead{jj@phys.uconn.edu}
\begin{abstract}
We develop a classical theoretical description for nonlinear many-body dynamics that incorporates the back-action of a continuous measurement process. The classical approach is compared with the exact quantum
solution in an example with an atomic Bose-Einstein condensate in a double-well potential where the atom numbers in both potential wells are monitored by light scattering. In the classical description the back-action of the measurements appears as diffusion of the relative phase of the condensates on each side of the trap. When the measurements are frequent enough to resolve the system dynamics, the system behaves classically. This happens even deep in the quantum regime, and demonstrates how classical physics emerges from quantum mechanics as a result of measurement back-action.
\end{abstract}


\maketitle
It is now possible to carry out experiments with systems comprised of a small and controllable number of atoms, and  eventually the interface between quantum mechanics and classical physics~\cite{Joos,Zurek,WAL85} must be confronted. Measurement back-action is an essential element of quantum mechanics, and may play a key role also in the transition from quantum mechanics to classical physics~\cite{LEG91,JAV96,Castin1997,Cirac1996}. In this paper we introduce a theoretical method to account for measurement back-action in nonlinear dynamics~\cite{SPI94,BRU97,Jacobs,Chaudhury,Scott,JAV08} classically, using a continuously monitored~\cite{RUO98} Bose-Einstein condensate (BEC) in a double-well trap as an explicit example. When the continuous measurements are strong enough so that the system dynamics can be resolved,
the classical description and a full quantum analysis agree well even deep within what might appear to be the domain of quantum mechanics.

\begin{figure}[b]
\begin{center}
\includegraphics[width=8cm]{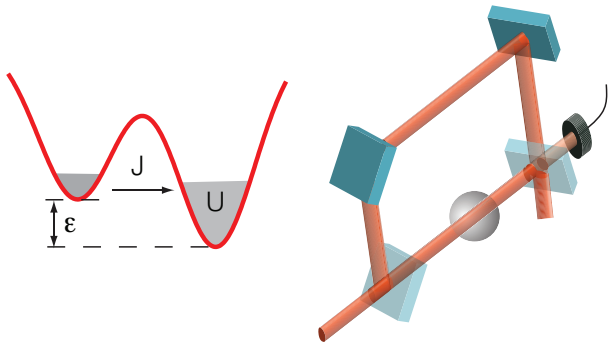}
\end{center}
\caption{
Left: Bose-Einstein condensate in a double-well potential with tunneling amplitude $J$, atom-atom interaction strength $U$, and energy difference per atom between the wells $\epsilon$ indicated.  Right: Basic idea of detection of the atoms in one trap. Since the detection light is far-off resonance, the dominant effect of the atoms is to alter the phase of the light passing through the sample by an amount proportional to atom number. The phase shift is then detected interferometrically. The amplitude of the light passing to the photodetector is proportional to the number of atoms, so that the rate of photon counts is proportional to the square of the atom number. The scheme is heavily simplified. For instance, mode matching of the light to the condensate is not considered.}
\label{SILLYFIGURE}
\end{figure}

We employ the minimal two-mode model~\cite{MIL97} to describe the Josephson oscillations of the atoms between the potential wells~\cite{JAV86,SME99,ALB05,LEV07,ZIB10}, see figure~\ref{SILLYFIGURE}. Given the hopping amplitude $J$ between the wells, the strength of atom-atom interaction inside a well $U$, and the imbalance in the energies of the atoms between the sides $\epsilon$, the Hamiltonian is
\beq
H=  -J(a^\dagger b + b^\dagger a)  +  U(a^\dagger a^\dagger  a a
 + b^\dagger b^\dagger bb)  +  \epsilon(a^\dagger a-b^\dagger b)\,.
\label{HAM}
\eeq
Here $a$ and $b$ are the boson operators for the atoms in the two wells, and we set $\hbar=1$.

The atom numbers in the two wells are monitored by coherent scattering of off-resonant  light; consult figure~\ref{SILLYFIGURE}. Counting of photons on two detectors, one for each potential well, constitutes the final measurement. In our quantum description we consider the atoms as an open system, and model the back-action of the measurements by adding a linear relaxation term to the equation of motion of the density operator of the atoms,
\beq
\dot\rho = -i[H,\rho] + {\cal L}\rho\,.
\label{MEQ}
\eeq
The most general relaxation term ${\cal L}\rho$  such that a density operator remains valid during the time evolution is of the  Lindblad form~\cite{LIN76,GAR04},
\beq
{\cal L}\rho = \sum_{i} \left[L_i \rho L^\dagger_i - \half (\rho L_i^\dagger L_i +L_i^\dagger L_i\rho )\right]\,,
\label{MEASURE}
\eeq
where $L_i$ are operators of the double-well system.
In the present model~\cite{RUO98} we have the Lindblad operators $L_a= \sqrt\Gamma\,a^\dagger a$ and $ L_b = \sqrt\Gamma\, b^\dagger b$, indicating that the scattering rates of photons from the two wells are $\Gamma \langle   (a^\dagger a )^2\rangle$ and $\Gamma \langle  ( b^\dagger b )^2 \rangle$, respectively. If the light is predominantly scattered into a narrow cone in the forward direction,
$\Gamma$ also specifies the strength of the measurements. It can be adjusted, for instance, by varying the intensity or the frequency of the probing light.

One could develop a parallel classical description using various phase-space representations such as the Wigner function for the ordinary~\cite{DRU93,WAL94,GAR04} or the SU(2)~\cite{NAR74,KOR77,KOR79} coherent states, or the number-phase Wigner representation~\cite{HUS11}. We choose the standard Wigner function because it is generally used in quantum optics and is not specifically tailored to the present system. We begin by rewriting the master equation~\eq{MEQ} equivalently in terms of the Wigner function $W(\alpha,\alpha^*,\beta,\beta^*;t)$~\cite{WAL94,GAR04}, where  $\alpha$ and $\beta$ are complex numbers corresponding to the operators $a$ and $b$. In the equation of motion of the Wigner function, $\alpha$, $\alpha^*$  and $\beta$, $\beta^*$ behave like canonical pairs with Poisson brackets such as $\{ \alpha,\alpha^*\}=-i$.  We then change to four real variables $N$, $\Phi$, $Z$ and $\varphi$ defined by
\beq
\alpha = \sqrt{\half N - Z}\, e^{i (\Phi-\half\varphi)},\,\beta =  \sqrt{\half N + Z}\, e^{i (\Phi+\half\varphi)}\,.
\eeq
This transformation between the variables is canonical, and results in Poisson brackets such as $\{Z,\varphi\}=1$. To the extent possible, we regard the Wigner function as the classical joint probability density for total atom number $N$, imbalance of atom numbers between the sides of the double well $Z$ ($Z\in[-N/2,N/2]$), overall phase of the  two BECs $\Phi$, and phase difference of the BECs between the two potential wells~$\varphi$. The global phase $\Phi$ may be dropped as it has no dynamics, nor any relevance to the dynamics of the other variables. We define $\chi=UN$, $z=2Z/N$ ($z\in[-1,1]$), and expand the equation of motion of the Wigner function up to and including the order ${\cal O}(N^{-1})$ in the formally large atom number $N$. The equation of motion
\beq
\partial_t W = \left[ -\frac{\partial {h}}{\partial \varphi}\,\partial_z
+ \frac{\partial h }{\partial z} \,\partial_\varphi +  \Gamma\,\partial_{\varphi\varphi} \right]W
\label{WFEQ}
\eeq
follows for the remaining distribution function $W(z,\varphi)$, with the effective one-particle Hamiltonian
\beq
 h(z,\varphi)=z^2\chi -2\epsilon z-2 J \sqrt{1-z^2} \cos (\varphi )\,.
 \label{ONEHAM}
 \eeq

Importantly, the quantum mechanical back-action of the measurements of the particle numbers is now accounted for by classical diffusion of the relative phase of the BECs  proportional to $\Gamma \partial_{\varphi\varphi}$. This is, perhaps, as expected: The relevant measurement result is the difference of the photon counting rates on the two detectors, and this is proportional to the expectation value of the operator
\beq
\Delta\hat N = (a^\dagger a)^2 - (b^\dagger b)^2 = (a^\dagger a + b^\dagger b)(a^\dagger a - b^\dagger b)\,.
\eeq
But the first factor in the final form is the conserved total atom number, so the measured operator is actually the difference of atom numbers between the sides of the trap. Given the (in quantum mechanics, qualitative) conjugate relation between number and phase, measurement of atom number difference should indeed increase the uncertainty of the relative phase.

We compare the quantum and classical evolutions in the case when the system is first prepared to the ground state at zero temperature in the presence of an energy imbalance, $\epsilon\ne0$, and the imbalance is suddenly removed by setting $\epsilon=0$. The atoms then start oscillating between the two sides of the potential well.

In our quantum analysis the evolution of the density operator according to~\eq{MEQ} is solved using quantum trajectory simulations~\cite{DAL92,DUM92,TIA92}. For every run of the simulation we have a state vector that undergoes `quantum jumps' at random times according to a precisely defined probabilistic law. As discussed in, say, Refs.~\cite{GAR04,WIS10}, here we also assume an ``efficient'' measurement process in which every scattered photon gets detected. Then a quantum trajectory simulation with its quantum jumps produces a faithful simulation of an individual experimental run, each quantum jump corresponding to a detection of a photon. Even though the master equation~\eq{MEQ} is not conditioned on the results of the measurements and as such describes an experiment in which the measurement results are discarded, the results {\em are\/} built into each quantum trajectory.

In a real experiment one would record a stream of photon counts, and the population imbalance $z(t)$ has to be inferred from them in some manner. For instance, if frequent photon counts merge into an essentially continuous photocurrent that varies on a time scale of the back-and-forth oscillations of the atoms, it would be sensible to regard the difference of the photocurrents coming from the two detectors as a  measure of the instantaneous population imbalance. However, for the sake of a clean illustration we analyze our quantum simulations differently: Given a quantum trajectory, a time dependent state vector, we use the expectations value of $\hat z = (a^\dagger a-b^\dagger b)/N$ to represent the value of the experimentally inferred population imbalance $z(t)$.  Evidently this works the better, the more numerous  the quantum jumps are on the time scale of the variation of $z(t)$.

In the classical analysis we take into account the quantum fluctuations in the initial state in the same way as is customary with the truncated Wigner approximation (TWA)~\cite{ISE06,BLA08,POL10,MAR10}. Thus, suppose equations~\eq{WFEQ} and~\eq{ONEHAM} were written in terms of the variables $Z$ and $\varphi$ instead of $z$ and $\varphi$. We expand the corresponding Hamiltonian ${\cal H}(Z,\varphi)$ to second order in $Z$ and $\varphi$ around the classical lowest-energy state in the presence of the initial energy imbalance $\epsilon\ne0$ between the sides. Since the classical variables $Z$ and $\varphi$ are a canonical pair, we may {requantize\/} by postulating the commutator $[Z,\varphi]=i$. ${\cal H}(Z,\varphi)$ thereby becomes the quantum Hamiltonian of a harmonic oscillator. The Wigner function corresponding to the ground state of this oscillator is used as the initial distribution of the variables $Z$ and $\varphi$, and leads to the initital condition for the distribution of the variables $z$ and $\varphi$
\beq
W_0(z,\varphi)  =  \frac{N}{2\pi}\exp\left[-\frac{N}{2}\left(\frac{(z-f)^2}{\xi} + \xi\varphi^2\right)\right].
\label{INDIST}
\eeq
We take $\chi>0$ and choose $\epsilon$ in such a way that the classical equilibrium has the phase $\varphi=0$ and a predetermined population imbalance $z=f$.
The parameter
\beq
\xi= \frac{1-f^2}{\sqrt{1+(1-f^2)^{3/2}\chi/J}}
\eeq
reflects the size of the ground state wave function of the underlying harmonic oscillator.

The evolution~\eq{WFEQ} with the initial Wigner function $W_0(z,\varphi)$ is solved by unravelling the dynamics into stochastic trajectories~\cite{ISE06,BLA08,GAR09,KOR77}.  If there were no phase diffusion in~\eq{WFEQ}, we would have an ensemble of trajectories $(z(t),\varphi(t))$ evolving under the effective Hamiltonian $h$, which makes the usual TWA. Given the diffusion, there exists a precise mathematical correspondence between~\eq{WFEQ} and the stochastic differential equations
\bea
dz &=& 2J\sqrt{1-z^2}\,\sin\varphi\,dt,\label{LEQ1}\\
d\varphi &=& -2z\left(\chi +J\frac{\cos\varphi}{\sqrt{1-z^2}}\right)\,dt+\sqrt{2\Gamma}\,dW(t)\,,\label{LEQ2}
\eea
where $dW(t)$ denotes a random increment of the Wiener process over the interval $dt$~\cite{GAR09}. The evolution of the distribution $W(z,\varphi;t)$ from the initial condition $W_0(z,\varphi)$ according to~\eq{WFEQ}   is obtained from an ensemble of stochastic trajectories. Specifically, the initial state of each trajectory $(z,\varphi)$ is sampled from the distribution  $W_0(z,\varphi)$, the dynamics of the trajectory $(z(t),\varphi(t))$ is determined from~\eq{LEQ1} and~\eq{LEQ2}, and the expectation value of any quantity over the distribution $W(z,\varphi;t)$ is correctly obtained by averaging the value of the quantity over a large number of such trajectories. The stochastic dynamics is  integrated  numerically using the Milstein algorithm~\cite{GAR09}.

We are now in a position to state precisely what ``classical description'' means in this context: unlike quantum mechanics, the analysis conforms to classical logic. The usual mean-field theory~\cite{SME99} is a classical description, as are the TWA and our present approach. The initial Wigner function $W_0$ is a valid classical probability distribution, even when it approximately incorporates quantum fluctuations of the atomic populations in  the initial state. Moreover, subsequent evolution according to~\eq{WFEQ} that also accounts for measurement back-action keeps the probability distribution  function valid at all future times. Correspondingly, we have been able to represent the complete solution in terms of classical stochastic trajectories. Such a trajectory exists objectively even without observations, but otherwise a similar correspondence between simulations and experiments prevails in the classical description as in quantum mechanics: each individual stochastic realization would be a representative outcome of an individual experiment.

\begin{figure}
\begin{center}
\includegraphics[width=8cm]{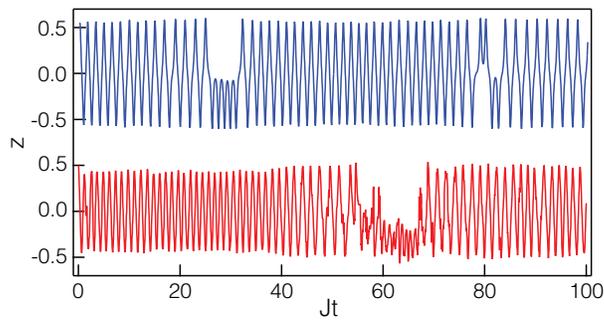}
\end{center}
\caption{An individual classical (upper trace, blue) and quantum (lower trace, red) trajectory. The parameters are $N=100$, $\chi/J = 10$, $f=0.5$, and $\Gamma=0.003$.}
\label{INDTRA}
\end{figure}

Figure~\ref{INDTRA} shows a classical (blue upper trace) and a quantum rendition (red lower trace)  of a single experiment.  In the mean-field theory, and for each individual TWA trajectory, one expects strictly periodic oscillations in the classical trace. The deviations from periodicity originate from the phase diffusion. On the other hand, the measurement strength $\Gamma/J=0.003$ is chosen such that each photon detector produces about ten clicks during a typical back-and-forth swing of the atoms, so that the quantum trace is a fair representation of a measurement record. We have deliberately picked these random examples so that qualitatively similar classical and quantum behavior with occasional trapping of the populations~\cite{SME99} in one or the other well is seen.

 \begin{figure}
\begin{center}
\includegraphics[width=8cm]{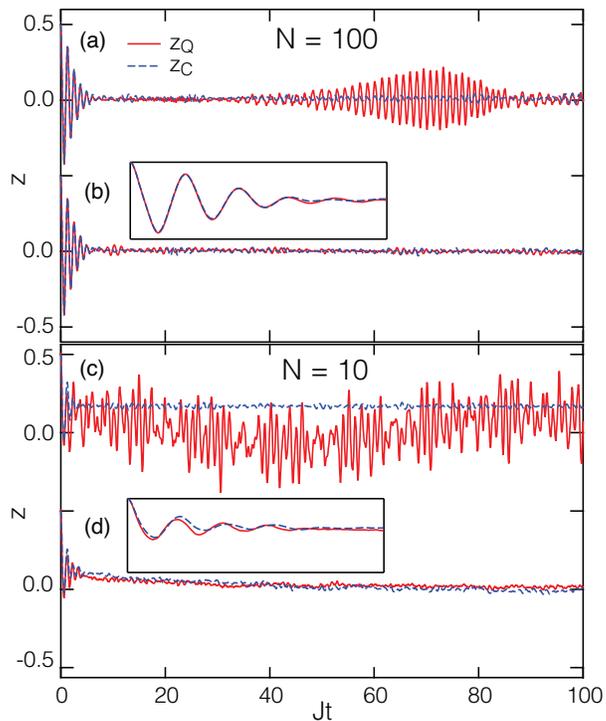}
\end{center}
\caption{Quantum (solid red lines) and classical (dashed blue lines) expectation values of the population imbalance $z$ as a function of time $t$, all averaged over $1,000$ trajectories. The upper half of the figure is for atom number $N=100$, the lower half for $N=10$. Graphs (a) and (c) are without continuous monitoring, $\Gamma/J=0$, graphs (b) and (d) correspond to the observation strengths $\Gamma/J=0.003$ and $\Gamma/J=0.3$ . The atom-atom interaction strength is $\chi/J = 10$, and the initial population imbalance is $f=0.5$. The insets demonstrate the quality of the quantum-classical agreement in the collapse region of traces (b) and (d).}
\label{QMVSCL}
\end{figure}

Nevertheless, even if we are able to model individual experiments both classically and quantum mechanically, these are stochastic stimulations. In order to construct expectation values of experimental observables for quantitative comparisons we have to average over many trajectories.
In figure~\ref{QMVSCL} we demonstrate simulation results for population imbalance when averaged over a thousand trajectories. The upper two panels are for $N=100$ atoms and the lower two for $N=10$, while the differences between the graphs (a)--(d) are in the measurement strength $\Gamma$.

Panel (a) for $N=100$ atoms plots the quantum and classical  expectation values of the population imbalance $z(t)$ with no measurements, $\Gamma=0$. The quantum result then represents the usual quantum average of the operator $\hat z= (a^\dagger a -b^\dagger b)/N$, and has nothing to do with the (nonexistent) continuous monitoring. The quantum and classical graphs both show a collapse up to $Jt\simeq 10$, but only the quantum expectation value has a revival at around $J t=70$.  In panel (b) we turn up the measurements strength to $\Gamma/J = 3\times10^{-3}$. Both detectors then report about ten photons during a characteristic oscillation time of the atoms between the traps, so that the oscillations of the atoms could be resolved by monitoring the light scattered off the atoms. The quantum expectation value of $\hat z$ should therefore be a passable representation of the population imbalance inferred from photon counts. The collapse part, quantum and classical, is almost unchanged by the measurements, but the quantum revival has disappeared. The quantum and classical analyses now agree very well.

We finally move on to the atoms number $N=10$. In panel (c) we show the population imbalances as a function of time for both the quantum theory and the classical theory without continuous monitoring, $\Gamma=0$. There is little in common with the two traces, and one would conclude that the classical description is totally off. The proximate reason is that with a decreasing atom number the quantum revivals start overlapping and produce irregular-looking behavior that is absent in the classical approach. It appears that we are deep in the domain of quantum mechanics. However, in panel (d) we have continuous measurements with the strength $\Gamma/J = 0.3$, which again gives about ten photon counts during a back-and forth oscillation of the atoms; and again, the classical and quantum results agree.

We have surveyed the behavior of the system for parameter values comparable to those in figures~\ref{INDTRA} and~\ref{QMVSCL}. In the absence of measurements, $\Gamma=0$, we have the usual TWA and the corresponding full quantum theory.  The TWA relies on a stochastic description and cannot fully incorporate quantum effects. The generic observation is that TWA agrees well with the exact solution for short times when stochastic field is highly occupied, but becomes progressively less accurate a model of quantum dynamics when the average mode population becomes smaller and the simulation time scales longer~\cite{ISE06,BLA08,POL10}. There are also examples, however, especially in lower dimensional systems in which cases the TWA simulations have been successful in providing qualitative descriptions of experimental findings even in the strongly fluctuating quantum limit~\cite{RUO95}.

In figure~\ref{QMVSCL}, in traces (a) with $N=100$ and $\Gamma=0$ we find a reasonable agreement between the TWA and quantum solutions up to the quantum revival around $Jt=70$, while in figure~\ref{QMVSCL}(c) with $N=10$ the two cases part ways after one back-and forth swing of the atoms at $Jt\simeq1$, a time that is hard to discern in the figure at all. Second, as the measurement strength is increased from zero, the changes in the quantum results start being qualitatively visible at times when there have been, on the average, a few counts on both detectors. The classical results also change with increasing $\Gamma$, but the changes are less conspicuous. For example, with $N=100$ the maximum amplitude of the quantum revival at $Jt\simeq70$ has decreased to half of its value in figure~\ref{QMVSCL}(a) for $\Gamma/J=2\times 10^{-5}$, at which point the average time between photon counts at each detector is about $20/J$. Let us next consider $\Gamma/J=3\times10^{-3}$, the value used in figure~\ref{QMVSCL}(b) for $N=100$. Since our definition of the parameter $\Gamma$ is such that the photon counting rate on the detectors is proportional to the square of the atom number, for $N=10$ this measurement strength is still modest, with the time between counts of about $10/J$.   A plot (not shown) would reveal that by $Jt=100$ the oscillations in the quantum result have lost 2/3 of their amplitude, but that the quantum and classical approaches still give quite different results.

Finally, as soon as the atom numbers are monitored vigorously enough so that the classical evolution in the form of the oscillations of the atoms between the two sides of the double-well trap can be resolved in detail in the scattered light, figures~\ref{QMVSCL}(b) and~(d), the quantum and classical descriptions give materially the same results. In round numbers, the time scale for classical evolution in our examples is $1/J$ and the scattering rate to both detectors is $N^2\Gamma/4$. The watershed criterion is therefore $N^2\Gamma/4J\sim1$, exceeded by a factor of about 10 in both graphs~\ref{QMVSCL}(b) and~(d). Moreover, and unlike in the usual case with the ordinary TWA, in our amended version of the TWA the quantum-classical agreement seems to persist for a long time, possibly as long as our numerical computations remain reliable.

The emergent classicality is a consequence of measurement back-action suppressing the subtle interference effects associated with quantum revivals that are inherent in the full quantum description~\cite{WAL85}. Remarkably, this occurs even deep in the domain of quantum mechanics with only 10 atoms, representing a far more dramatic transition to classicality than, e.g., measurement-induced relative phase between condensates~\cite{JAV96,Castin1997,Cirac1996} or nonlinear instabilities~\cite{Jacobs,Scott,JAV08,JAV10}.
The classical description  at this level of accuracy is only possible because we have included classical phase diffusion to account for quantum mechanical measurement back-action. To wit, the only difference between the classical predictions in graphs~\ref{QMVSCL}(c) and~(d) is the amount of phase diffusion, yet they are clearly different.

The main technical achievement of this paper is that we have introduced a theoretical method for a continuously monitored quantum system that handles both quantum fluctuations and measurement back-action classically. Now, the computational resources required to solve the time evolution from quantum mechanics tend to increase very fast with the system size. We therefore  hope to inspire practical algorithms to analyze  time evolution in situations where quantum effects are small but not negligible.

\ack This work is supported in part by NSF, Grant No. PHY-0967644, the Leverhulme Trust, and the EPSRC.

\section*{References}
\bibliographystyle{unsrt}

\end{document}